\documentclass[a4paper,11pt]{article}
\usepackage{pos}
\usepackage{graphicx}
\usepackage{float}

\title{Galactic calibration and its long-term stability for the Auger Engineering Radio Array}
\ShortTitle{Galactic calibration of AERA}

\author[a]{D. C dos Santos}

\author[b, \dag]{ for the Pierre Auger Collaboration\notes{\note{Full 
author list at \url{http://www.auger.org/archive/authors_2024_06.html}.}}}

\affiliation[a]{Instituto de Física, Universidade Federal do Rio de Janeiro,\\
  Rio de Janeiro, Caixa postal 68528, Rio de Janeiro, Brazil}

\affiliation[b]{Observatorio Pierre Auger, Av. San Mart\'in Norte 304,
5613 Malarg\"ue, Argentina.}

\emailAdd{spokespersons@auger.org}

\abstract{The Auger Engineering Radio Array (AERA), part of the Pierre Auger Observatory, is a facility designed to detect radio emissions from extensive air showers at high energies. Consisting of 153 autonomous radio-detector stations spread over 17 km$^2$, it detects radio waves in the frequency range of 30 to 80 MHz.  Accurate characterization of the detector response is essential for correct data interpretation, previously achieved through laboratory measurements of the analog chain and measurements of the antenna's directional response. In this study, we perform an absolute calibration using the continuously monitored sidereal modulation of the diffuse Galactic radio emission. Calibration is done by comparing the average spectra recorded by the stations with seven different models of the full radio sky propagated through the system response, including antennas, filters, and amplifiers. The Galactic calibration is in good agreement with the original laboratory calibration. In addition, we analyze the time-dependence of the calibration constants over a period of seven years. No aging effects are observed in AERA stations over a timescale of a decade, which shows that radio detectors could help monitor possible aging effects of other detector systems during long-term operations and highlight their importance in determining an absolute cosmic-ray energy scale.}

\FullConference{%
  10th International Workshop on Acoustic and Radio EeV Neutrino Detection Activities - ARENA2024\\
  11-14 June 2024\\
  The Kavli Institute for Cosmological Physics, Chicago, IL, USA}

\begin{document}
\maketitle

\section{Introduction}
\label{sec:intro}

\hspace{0.5cm}

The Auger Engineering Radio Array (AERA) \cite{AERA} is a facility of the Pierre Auger Observatory, specifically designed to capture radio emissions generated by high energy extensive air showers. It comprises 153 independent radio-detection stations distributed across 17 km$^2$, operating within the 30 to 80 MHz frequency range. Traditional calibration methods for AERA, involving laboratory measurements and drone-based, face challenges in accuracy and long-term applicability. In this study, we introduce an absolute Galactic calibration for AERA antennas, comparing recorded spectra with radio sky models to obtain calibration constants. We also analyze the stability of these constants from 2014 to 2020, revealing no significant aging effects in the antennas. This method not only improves data interpretation but also provides a reliable means for long-term monitoring of detector performance.

The proceeding is organized as follows: Section \ref{sec:dataset} details the detector system and the data set used for calibration.  Section \ref{sec:datacleaning} describes the data cleaning process, including the preprocessing steps to mitigate noise and ensure data quality. Section \ref{sec:radioskymodel} describes the radio sky models employed in the study, while Section \ref{sec:calibration} provides an explanation of the methodology for absolute Galactic calibration and presents the results. Section \ref{sec:aging} discusses the analysis of calibration constants over time. Finally, Section \ref{conclusions} summarizes the conclusions of the study.

\section{Detector system and data set}
\label{sec:dataset}

\hspace{0.5cm} As an engineering array, AERA has involved the development and testing of various antenna types, electronics, and trigger systems over time. This study utilizes data from Butterfly antennas and Logarithmic-Periodic Dipole Antennas (LPDA) \cite{AERA}. The LPDA antennas, employing the logarithmic periodic dipole principle, incorporate a series of half-wave dipoles with progressively increasing lengths to ensure consistent radiation resistance across a broad frequency spectrum. In contrast, Butterfly antennas, known as "bow tie" models, feature a much simpler design that consists of two triangular arms.

The absolute Galactic calibration is performed by using periodically triggered traces measured with 52 Butterfly antennas (2014-2020), 23 Butterfly antennas (2016-2020), and 14 LPDA antennas (2017-2020). The data acquisition involved periodically triggered traces every 100 seconds, with corrections applied for temperature-dependent gain variations in the Filter Amplifier (FA) and Low Noise Amplifier (LNA). The time series of each trace has a time duration of $5.7 \mu$s. This time series is then transformed into the frequency domain with a resolution of $\Delta \nu = 0.175$ MHz, covering the 30-80 MHz frequency range. For Galactic calibration purposes, we compute the power within each frequency band $\nu$, which has a width of $\delta \nu = 1$ MHz, by using

\begin{equation}
\label{eq:power}
    P_\nu = \frac{2}{T}\sum_{k=\nu - \delta \nu/2}^{\nu+\delta \nu/2}\frac{|V(k)|^2}{Z_L} \Delta \nu,
\end{equation}
where $T$ is the length of the trace, $V(k)$ is the  measured spectral voltage at frequency $k$ and $Z_L= 50 ~ \Omega$ is the antenna impedance. The factor of 2 arises from utilizing only half of the Fast Fourier Transform spectra.  Since all stations are located in approximately the same geographic region, the variation of the Galactic signal can be assumed to be the same for all antennas. However, the background signal includes both Galactic and other noise components, whose intensity varies by source. Therefore, data preprocessing to mitigate this noise is essential before performing Galactic calibration.

\section{Data cleaning}
\label{sec:datacleaning}
\begin{figure}[H]
        \centering
                \includegraphics[scale=0.17]{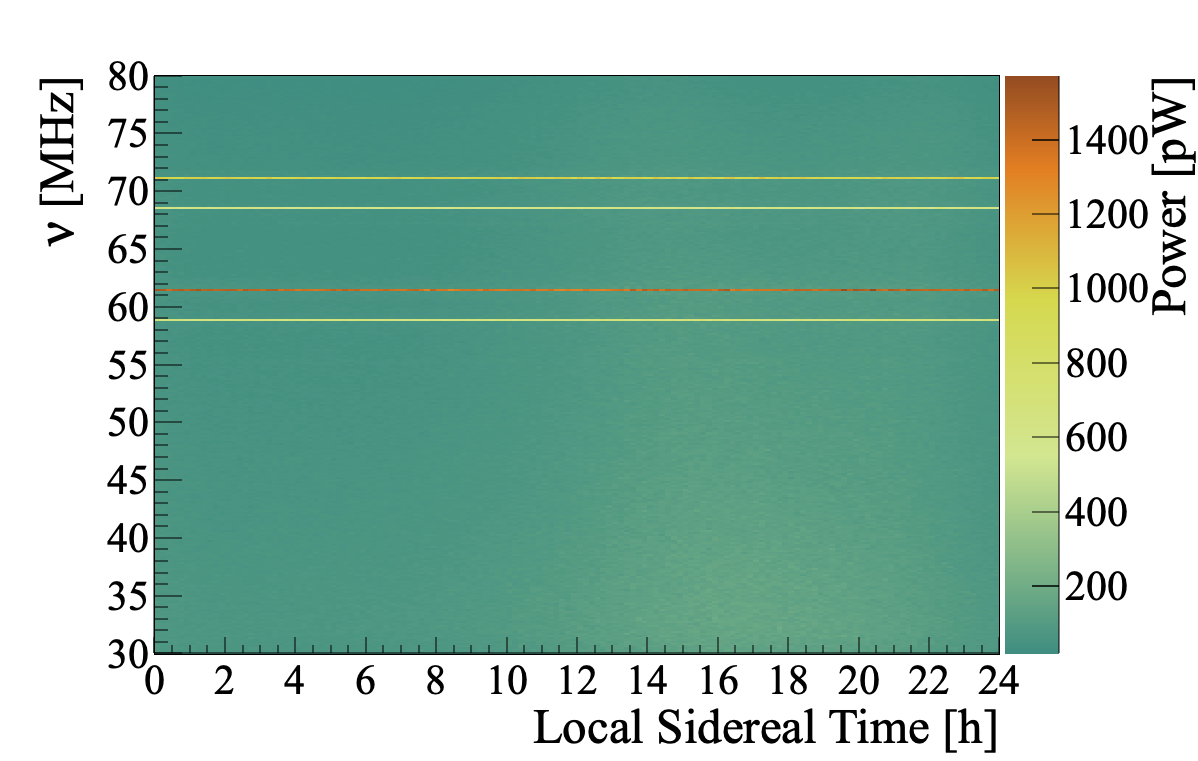} \quad
                \includegraphics[scale=0.17]{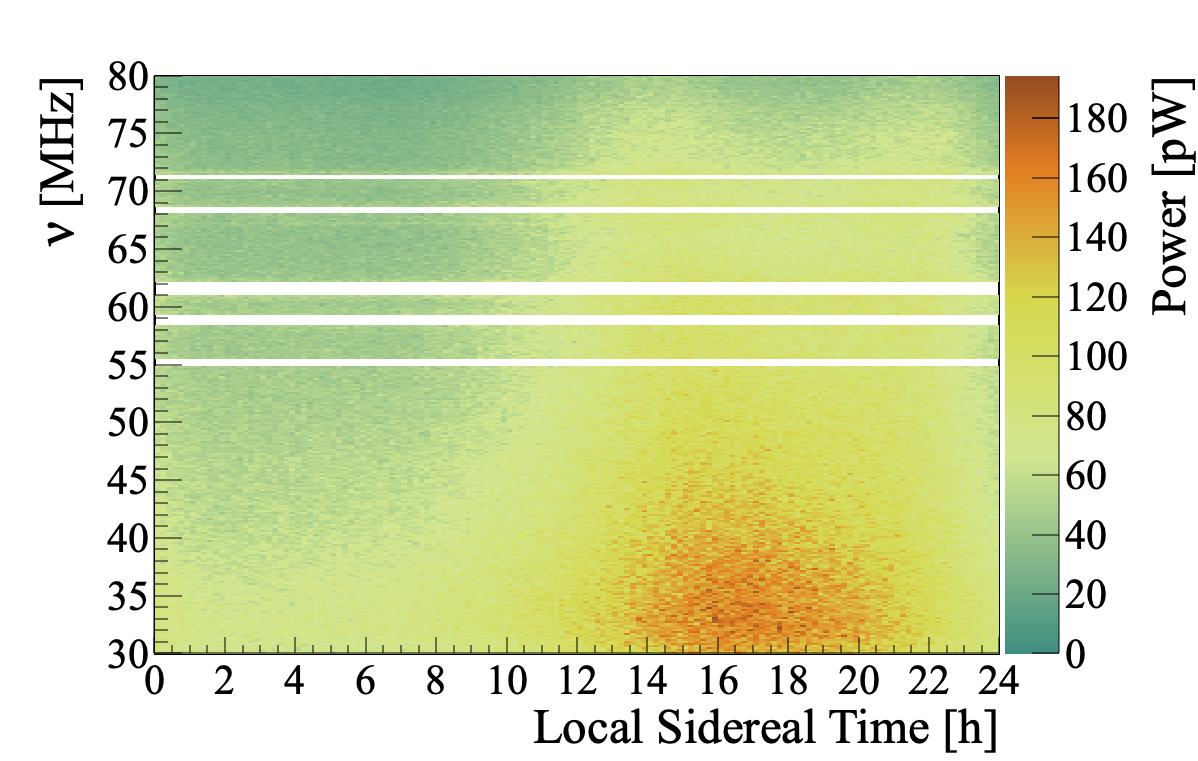} 
        \caption{\footnotesize{Dynamic average frequency spectrum as a function of LST for North-South channel for one specific antenna during February 2019. The left panel presents the results before removal of narrowband RFI (horizontal lines) while the right ones show that the Galactic signal variation becomes evident after RFI removal.}}
                \label{Fig:Spectrum_EW_NS}
 \end{figure}

\hspace{0.5cm} The passage of the Galaxy above the Auger site (69.3$^\circ$W, 35.3$^\circ$S) results in a Galactic modulation of the radio signal intensity as a function of Local Sidereal Time (LST). The periodic data that capture this modulation also contain noise that contaminates the measurements, including cosmic ray signals, Radio Frequency Interference (RFI) from external sources, and internal electronic noise. RFI is categorized as broadband or narrowband, with narrowband noise coming from continuous emissions at specific frequencies, such as the AERA beacons \cite{beacon} and TV line.  The amplitude of these noises makes it difficult to detect the passage of the Galaxy center in the dynamic spectrum of frequencies. To address this issue, the affected frequency bands are identified and replaced with interpolated values. Figure \ref{Fig:Spectrum_EW_NS} shows the dynamic spectrum for an antenna in February 2019, where Galactic modulation becomes clearer after noise suppression.

On the other hand, broadband RFIs, characterized by transient radio pulses, also contaminate the data.  To mitigate the impact of this noise, as well as cosmic-ray signals, we implemented a time-dependent threshold method in LST to discard undesired traces. This process begins by calculating the average spectral density of each trace according to

\begin{equation}
\label{eq:averageSpectralDensity}
     I = \frac{1}{n}\sqrt{\sum_{i=1}^{n}A^{2}(\nu_i)},
\end{equation}where $A(\nu_i)$ is the signal amplitude in frequency bin $i$ and $n$ is the total number of bins.

The threshold is established by performing a Gaussian fit on the $I$ values for each 20-minute LST bin, determining a threshold $I^{\mathrm{th}}_{\mathrm{LST}}$ at 3$\sigma$. Traces with spectral densities exceeding this threshold are discarded. A smoothing step is then applied to avoid abrupt variations between bins.

\section{Radio sky model}
\label{sec:radioskymodel}

\begin{figure}[H]
        \centering
            \includegraphics[scale=0.24]{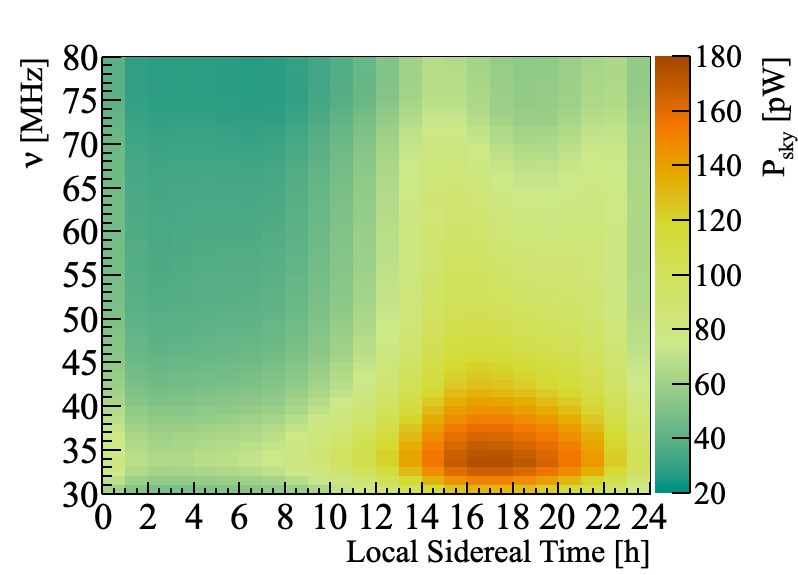} \quad
                \includegraphics[scale=0.24]{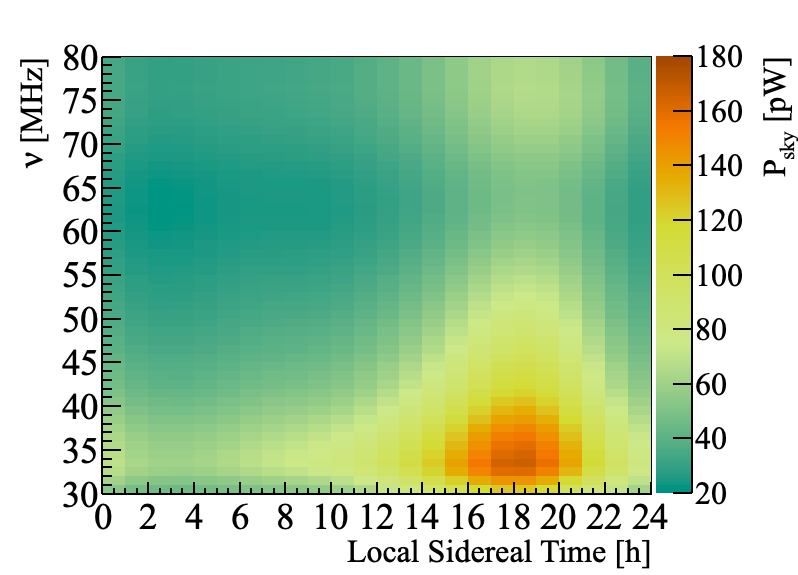} \quad
        \caption{\footnotesize{The left (right) panel shows the expected power $P_{\rm{sky}}$ to be received from the sky considering the LFmap model and the simulated VEL of Butterfly antennas as a function of LST and frequency for the North-South (East-West) channel. }}
                \label{Fig:p_model}
 \end{figure}

\hspace{0.5cm} The background radio signal received on Earth varies across different regions of the sky, and can be conveniently characterized by its equivalent brightness temperature, which is used to create radio sky maps. In the frequency range of AERA, the background signal is primarily dominated by Galactic emission. The total power expected to be received by the antenna can be expressed as

\begin{equation}\label{eq:model}
    P_{\mathrm{sky}}(t,\nu) = \frac{Z_0}{Z_\mathrm{L}}\frac{k_\mathrm{B}}{c^2} \int_{\Omega}\nu^2 T_{\mathrm{sky}}(\nu,\alpha,\delta)|H_e(\nu,\alpha,\delta)|^2 d\Omega,
\end{equation}where $k_\mathrm{B}$ is the Boltzmann constant, $c$ is the speed of light, $Z_{0}$ is the impedance of free space given by 120$\pi$ $\Omega$ and $Z_\mathrm{L}$ is the antenna impedance. $H_{e}(\nu,\theta,\phi)$ represents the Vector Effective Length (VEL), which is a 2-dimensional vector with complex components that describes the directional response of the antenna based on the frequency of an incoming radio signal from the direction $(\theta,\phi)$ \cite{AERA}. $T_{\mathrm{sky}}(\nu,\alpha,\delta)$ represents the sky temperature at a given frequency $\nu$, as a function of right ascension $\alpha$ and declination $\delta$. Several models exist to characterize sky radio emission, each using different techniques to map the radio-frequency sky across various frequencies, with some focusing on interpolation based on reference maps and others incorporating physical mechanisms of radio emission. Examples of these models include LFmap \cite{lfmap}, GSM 2008 \cite{GSM}, GSM 2016 \cite{GSM16}, LFSM \cite{LFSM}, GMOSS \cite{GMOSS}, SSM \cite{SSM}, and ULSA \cite{ULSA}. A systematic uncertainty estimate for predicting Galactic emission from sky models was derived in \cite{busken2022}. As an example, Fig. \ref{Fig:p_model} shows the expected power $P_{\mathrm{sky}}(t,\nu)$ received from the sky using the LFmap model and the simulated VEL for Butterfly antennas, presented in 1 MHz frequency bins and 1-hour LST bins for both polarizations.

\section{Calibration method and results}
\label{sec:calibration}

\hspace{0.5cm} For an accurate interpretation of AERA data, the antennas must be carefully calibrated, and a comprehensive understanding of the entire signal chain—including the antenna, amplifiers, filters, and digitizer—is essential to minimize measurement uncertainties. To accomplish this, we implemented a calibration approach inspired by the technique used in the LOFAR experiment \cite{lofar}. This method involves convolving the sky-emitted power with the gains and noise within the antenna signal chain, including any external environmental noise. This way, the model of the power emitted by the sky and propagated by the antenna can be described by
\begin{equation}\label{eq:model_final}
    P_{\mathrm{model}}(t,\nu) = P_{\mathrm{sky}}(t,\nu)\rm{G}_{\mathrm{ant}}(\nu)\rm{G}_{\mathrm{RCU}}(\nu){\rm C}_0^2(\nu) + \rm{N}_{\mathrm{tot}}(\nu),
\end{equation}in which $\rm{G}_{\mathrm{ant}}(\nu)$ and $\rm{G}_{\mathrm{RCU}}(\nu)$ are, respectively, the gains of the  LNA and of the Receiver Unit (RCU), where the signal is subjected to a bandpass filter, amplified and digitized. The free parameters of the model are the  the calibration constant $\rm{C}_0(\nu)$ and the total noise $\rm{N}_{\mathrm{tot}}(\nu)$, which is a combination of intrinsic electronic thermal noise and environmental noise.

\begin{figure}[h!]
        \centering
        \includegraphics[scale=0.51]{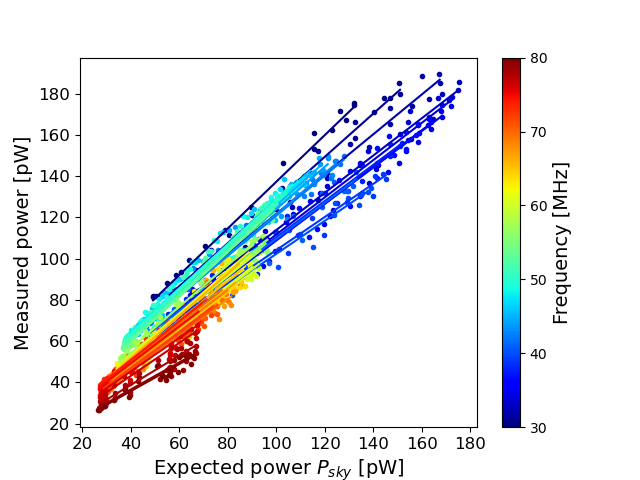}
        \includegraphics[scale=0.44]{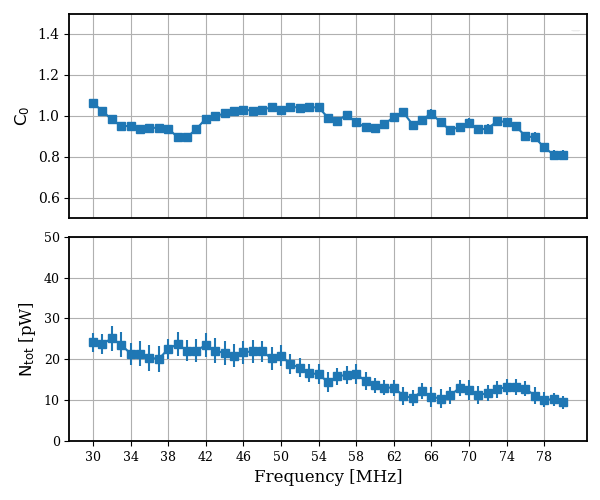}
        \caption{\footnotesize{Left panel: graph of the measured power versus the expected power, as well as the resulting linear fits obtained for each frequency bin of 1 MHz with respect to the North-South channel of one specific  Butterfly antenna by using periodic traces collected during January 2019. Right panels: calibration constants $\rm{C}_0$ (top panel) and total noise (bottom panel) obtained from the fit. 
}}
                \label{Fig:Fit_27}
 \end{figure}

To perform the calibration, we compare the measured signal at the antenna (Eq. \ref{eq:power}) with the expected signal (Eq. \ref{eq:model_final}). In other words, each frequency and LST bin of the measured power, as shown on the right side of Fig. \ref{Fig:Spectrum_EW_NS} for the North-South channel, is matched with the corresponding bin of the expected power, illustrated in the right panel of Fig. \ref{Fig:p_model}. A linear fit is then applied to each frequency band to determine the values of $\rm{C}_0^2(\nu)$ and $\rm{N}_{\rm{tot}}(\nu)$. The left panel of Fig. \ref{Fig:Fit_27} provides an example of the fit for the North-South channel of a specific antenna in January 2019, with the best fit parameters for each frequency range shown in the right panel.

\begin{figure}[H]
        \centering
                \includegraphics[scale=0.22]{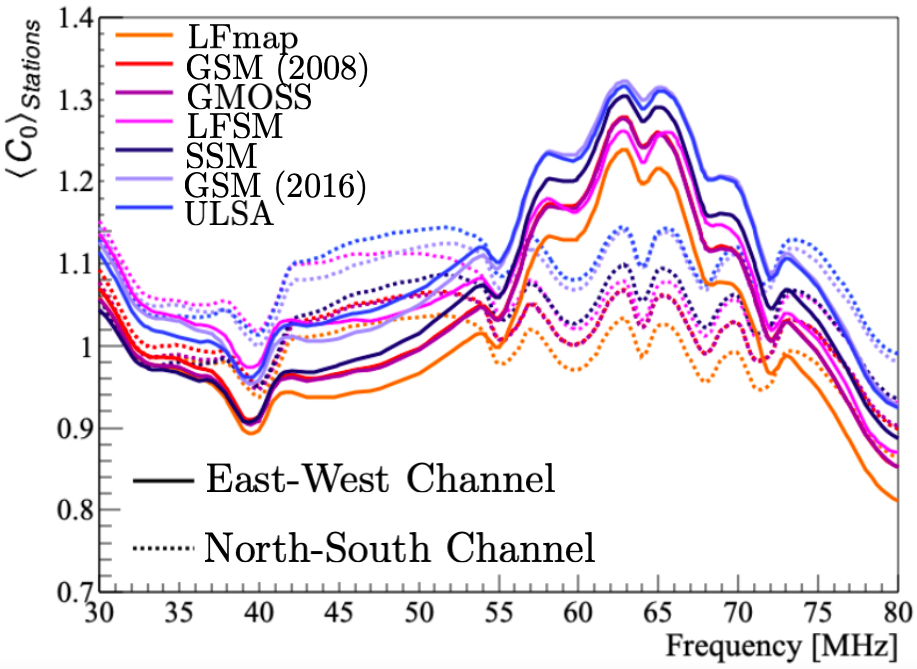} \quad
                \includegraphics[scale=0.22]{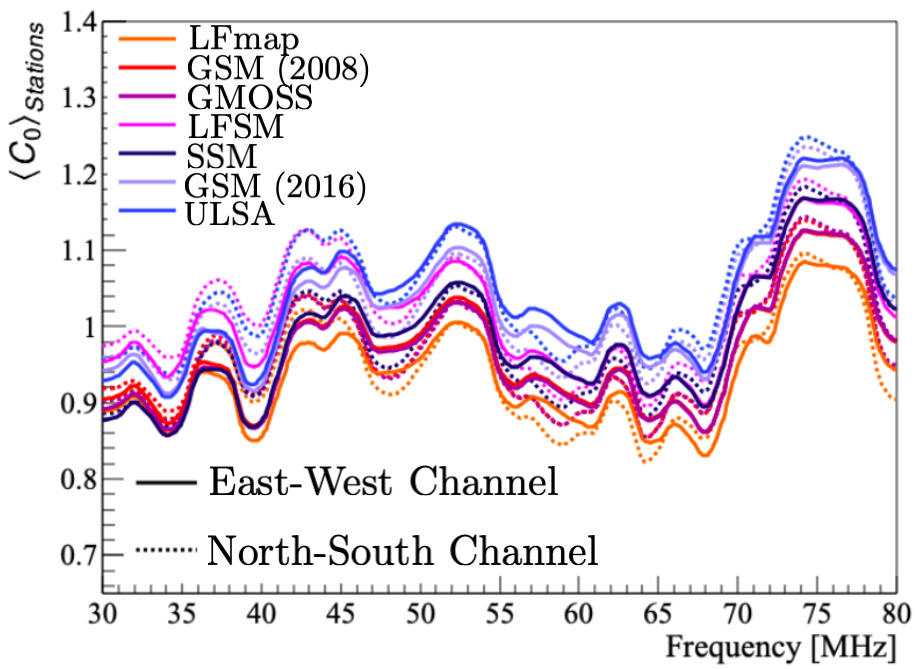} \quad
        \caption{\footnotesize{Average of calibration constants as a function of frequency for all Butterfly antennas (left panel) and LPDA antennas (right panel) for each sky temperature model, sampled monthly throughout the entire data period used in the quality selection. The North-South channel (dashed lines) and East-West channel (solid lines) exhibit similar behavior for LPDAs. However, for Butterfly antennas, a disparity is observed at higher frequencies due to directional response asymmetry.}}
                \label{Fig:results_constants}
 \end{figure}

 Figure \ref{Fig:results_constants} presents the average calibration constants across all antennas for each frequency band, with different sky models represented by various color lines. The left panel shows results for Butterfly antennas, while the right panel is for LPDA antennas. Different polarizations are distinguished by line styles, with LPDA antennas showing similar behavior between the North-South (dashed) and East-West (solid) channels. In contrast, Butterfly antennas display a discrepancy around 65 MHz, likely due to directional response asymmetry caused by an electronic box mounted on the main pole aligned with the East-West arm. An estimate of the impact of calibration constants on cosmic ray energy uncertainty can be obtained by considering the average calibration constant across all models. This average is defined $\left \langle \mathbb{C}_0 \right \rangle_{\rm{model}} = \frac{1}{N_\nu}\sum_{\nu} \left \langle \rm{C}_0(\nu) \right \rangle_{\rm{model}}$, where $\left \langle \rm{C}_0(\nu) \right \rangle_{\rm{model}} \equiv \frac{1}{N_{\rm{model}}}\sum_{i} {\rm C_{0,i}(\nu)}$. Here, $N_{\rm{model}} = 7$ represents the number of sky models used, and ${\rm{C}_{0,i}(\nu)}$ is the calibration constant obtained from each sky model $i$ at frequency $\nu$.

As shown in Fig. \ref{Fig:results_constants}, the LFmap and ULSA models provide the smallest and largest calibration constant values, respectively. Consequently, the systematic uncertainty, estimated at approximately 6\%, arises from discrepancies among these sky models, calculated as the mean deviation between the averaged model distribution and the LFmap and ULSA models. Statistical uncertainties, around 5\%, are estimated as the RMS of the averaged model distribution. The final estimator $\widehat{\mathbb{C}}_0$ for the calibration constants, summarized in Table \ref{tab:results_c0}, is derived from the average of the $\left \langle \mathbb{C}_0 \right \rangle_{\rm{model}}$ distribution. Since cosmic-ray energy is proportional to the square root of radiation energy \cite{AERA_energy} (which scales with $\widehat{\mathbb{C}}_0^2$), cosmic-ray energy is therefore proportional to $\widehat{\mathbb{C}}_0$. As a result, the impact of the systematic uncertainty on the absolute calibration of the radio cosmic-ray energy scale remains at the 6\% level.

\begin{table}[h]
\centering
\small
\begin{tabular}{lc}
\hline
\textbf{Station (channel)} & $\widehat{\mathbb{C}}_0$ $\boldsymbol{\pm \sigma_{\text{stat}} \pm \sigma_{\text{syst}}}$ \\
\hline
Butterfly (East-West) & $1.08 \pm 0.05 \pm 0.05$ \\
Butterfly (North-South) & $1.04 \pm 0.04 \pm 0.06$ \\
LPDA (East-West) & $1.01 \pm 0.07 \pm 0.06$ \\
LPDA (North-South) & $1.01 \pm 0.04 \pm 0.06$ \\
\hline
\end{tabular}
\caption{Calibration constant results obtained for both channels of Butterfly and LPDA antennas.}
\label{tab:results_c0}
\end{table}

\section{Study of the calibration constant as a function of time}
\label{sec:aging}
\begin{figure}[H]
        \centering
                \includegraphics[scale=0.35]{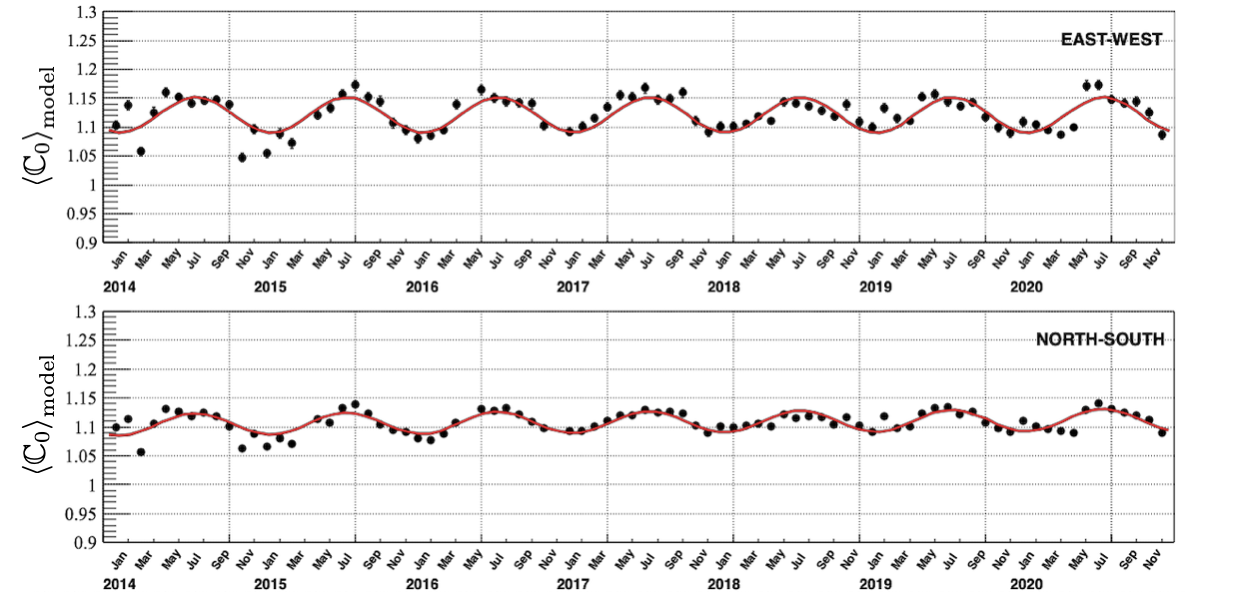}
        \caption{\footnotesize{Calibration constants obtained for both channels of antenna Id:33 from 2014 to 2020. The corresponding
cosine+linear fit is represented by the red curve.}}
                \label{Fig:C0_time}
 \end{figure}

\hspace{0.5cm} We analyzed the time behavior of calibration constants $\left \langle \mathbb{C}_0 \right \rangle_{\rm{model}}$ values over the entire data collection period for each antenna. Figure \ref{Fig:C0_time} shows these values for a specific Butterfly antenna from 2014 to 2020. Notice that even after correcting for the temperature-dependent gain variations in the amplifiers of the signal chain, a residual seasonal modulation of unknown origin is still observed.  A possible source might be the differences between our ambient temperature sensor and the actual temperature along the signal chain in the detector housing (due to cooling and heating). However, the effect of solar irradiation, as a proxy for the heating/cooling cycle, did not reveal a clear correlation. To account for this modulation, the calibration constant over time is modeled as \mbox{$\left \langle \mathbb{C}_0(t') \right \rangle_{\rm{model}} = A\cos{(\frac{\pi}{6}t' + \phi)} + at' + b$}, where $t'$ represents the time in months. The parameters $A$ and $\phi$ denote the magnitude and phase of the observed seasonal modulation, respectively, centered around the baseline value $b$ (in the absence of aging). The slope parameter $a$ is particularly important as it indicates the aging of the station per month.  The red curve in Fig. \ref{Fig:C0_time} illustrates the fit of the calibration constants over time, which was performed for all antennas in this study. It is worth mentioning that the fit does not perfectly capture the seasonal modulation because there are fluctuations in the $\left \langle \mathbb{C}_0 \right \rangle_{\rm{model}}$ values for all antennas during certain periods. These fluctuations impact the aging coefficient, especially when significant fluctuations occur at the start or end of the data collection period. Therefore, it is crucial to not confuse genuine aging effects with transient fluctuations that happen to occur during later or earlier years. To mitigate these fluctuations, we conducted mock simulations, generating random Gaussian-distributed calibration constants with mean and Root Mean Square (RMS) equal to the measured $\left \langle \mathbb{C}_0 \right \rangle_{\rm{model}}$ value and its corresponding uncertainty for each antenna and month, and fitting these simulated values over time.  We then systematically shuffled the years for all antennas and fitted the simulated data over time. The average mock aging coefficients were expected to converge to zero, with the RMS of the distribution providing the uncertainty estimate for the aging parameter.

\begin{figure}[H]
        \centering
                \includegraphics[scale=0.24]{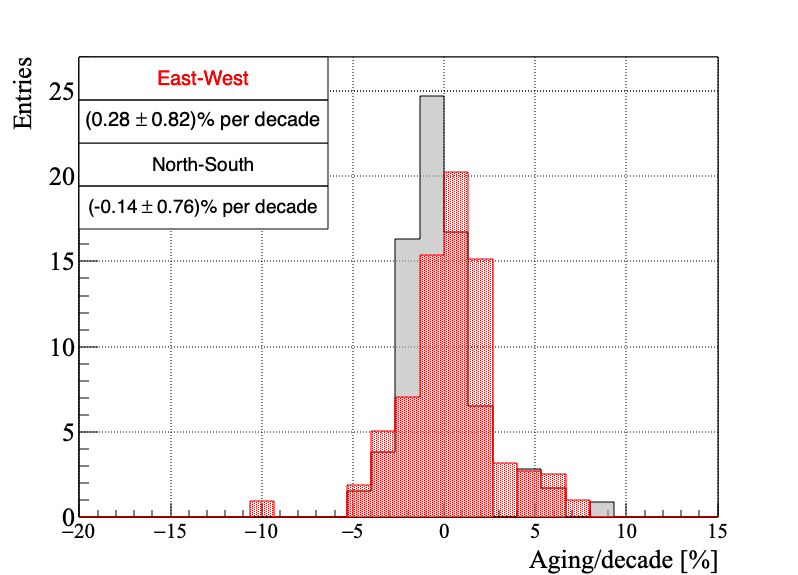} \quad 
                \includegraphics[scale=0.24]{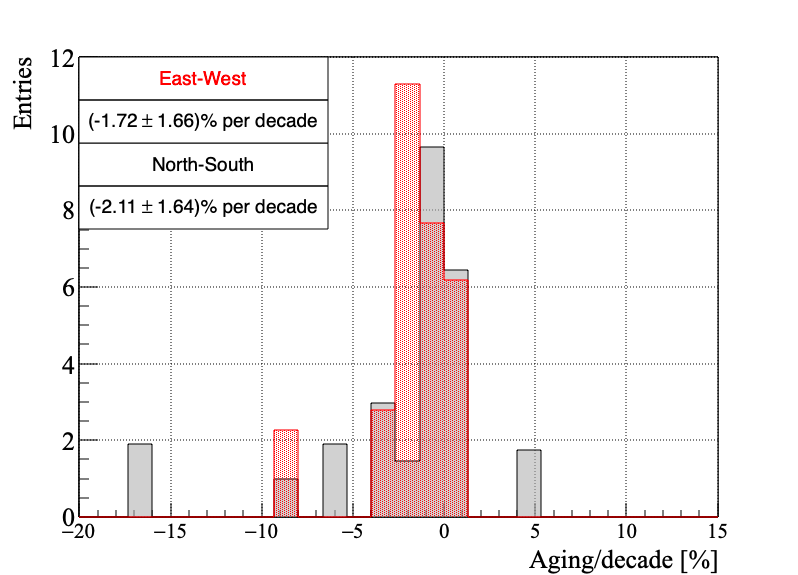} \quad 
        \caption{\footnotesize{Distribution of the angular coefficients $a$ (interpreted as the aging factor per month) in percentage per decade, obtained from each antenna and channel. Left: Butterfly stations. Right: LPDA stations.}}
                \label{Fig:aging_results}
 \end{figure}

Figure \ref{Fig:aging_results} shows the distribution of aging factors per decade for both Butterfly and LPDA antennas. We determined an aging factor of (0.28 $\pm$ 0.82)\% and ($-$0.14 $\pm$ 0.76)\% per decade for the East-West and North-South channels, respectively, for the Butterfly stations. For the LPDA antennas, the aging factor was found to be ($-$1.72 $\pm$ 1.66)\%  and ($-$2.11 $\pm$ 1.64)\% per decade for the East-West and North-South channels, respectively. Notice that the results from the LPDA are more prone to large fluctuations because the statistics are lower (only 14 antennas) and a shorter data collection period (4 years) was used. When combining all antenna types and channels,  we obtain an aging factor of $a = (-0.32 \pm 0.51) \%$ per decade. These results suggest that there is no significant aging effect observed in the AERA antennas.

\section{Conclusions}
\label{conclusions}
In this study, we performed an absolute frequency-dependent Galactic calibration of AERA stations. Using seven different radio sky models, we determined the average calibration constant $\widehat{\mathbb{C}}_0$ for each channel. The results show that the calibration constants are consistent with unity within uncertainties, indicating a good agreement with the original calibration process, which included laboratory measurements of the analog chain, and simulations to ascertain the directional response of the antennas. Additionally, we investigated the behavior of the calibration constants over time for a period of seven years. The calibration constants remain stable with negligible aging effects, consistently close to zero within the uncertainties. This highlights the effectiveness of radio detectors for long-term monitoring of aging effects in other detectors. The results are important for the Pierre Auger Observatory upgrade \cite{AugerPrime} and for the radio detection technique in determining an absolute energy scale \cite{ProceedingMax} for cosmic rays.


\begin{thebibliography}{99}
\small
\setlength{\itemsep}{1pt}

\bibitem{AERA}P.Abreu \textit{et al.}, JINST \textbf{7}  (2012) 10011



\bibitem{beacon}Schroder, Frank. Springer Science \& Business Media (2012) 1


\bibitem{lfmap}E. Polisensky, Long Wavelength Array Memo Series \textbf{111} (2007) 515

\bibitem{GSM}A. O Costa \textit{et al.}, MNRAS \textbf{388} (2008) 247


\bibitem{GSM16}H. Zheng \textit{et al.}, MNRAS \textbf{464} (2016) 3486


\bibitem{LFSM}J. Dowell \textit{et al.}, MNRAS \textbf{469} (2017) 4537


\bibitem{GMOSS} M. S. Rao \textit{et al.}, AJ \textbf{153} (2017) 26

\bibitem{SSM} Q. Huang, Science China Physics, Mechanics, and Astronomy \textbf{62} (2019) 989511


\bibitem{ULSA}Y. Cong \textit{et al.}, ApJ \textbf{914} (2021) 128


\bibitem{busken2022} M. Büsken, A\&A \textbf{679} (2023) A50


\bibitem{lofar} K. Mulrey, Astroparticle physics \textbf{111} (2019) 1


\bibitem{AERA_energy} P.Abreu \textit{et al.}, Phys. Rev. D \textbf{93} (2016) 122005


\bibitem{AugerPrime}P.Abreu \textit{et al.}, arXiv: 1604.03637


\bibitem{ProceedingMax}M. Büsken \textit{et al.}  \textit{PoS} \textbf{ARENA2024} (2024) 035

\end{thebibliography}

\end{document}